\documentstyle[epsfig]{mn}

\title[Asymmetric model for FRII sources]
	{An asymmetric relativistic model for classical double radio sources}
      
\author[T.\,G.\,Arshakian \& M.\,S.\,Longair]
       {T.\,G.\,Arshakian\thanks{On leave from Byurakan Astrophysical Observatory,        Byurakan 378433, Armenia}
        and M.\,S.\,Longair\\
        Cavendish Astrophysics Group, Cavendish Laboratory, Madingley Road, Cambridge, CB3 0HE}

\date{ }

\pagerange{\pageref{firstpage}--\pageref{lastpage}}

\pubyear{1999}

\begin{document}

\label{firstpage}

\maketitle

\begin{abstract}

There is substantial observational evidence against the symmetric relativistic model of FRII radio sources. An asymmetric relativistic model is proposed which takes account of both relativistic effects and intrinsic/environmental asymmetries to explain the structural asymmetries of their radio lobes. A key parameter of the model is the jet-side of the double sources, which is estimated for 80\% of the FRII sources in the 3CRR complete sample. Statistical analyses of the properties of these sources show that the asymmetric model is in agreement with a wide range of observational data, and that the relativistic and intrinsic asymmetry effects are of comparable importance. Intrinsic/environmental asymmetry effects are more important at high radio luminosities and small physical scales. The mean translational speed of the lobes is found to be $\overline{v}_{\rm lobe}=(0.11 \pm 0.013)\,c$, consistent with the speeds found from spectral ageing arguments. According to a Gaussian model, the standard deviation of the distribution of $v_{\rm lobe}$ is $\sigma_{v_{\rm l}} = 0.04c$.  The results are in agreement with an orientation-based unification scheme in which the critical angle separating the radio galaxies from quasars is about $45^{\circ}$.

\end{abstract}


\section{Introduction}
The classical double FRII radio sources are among the most luminous extragalactic radio sources (Fanaroff \& Riley 1974). They are characterised by two steep spectrum radio lobes, symmetrically disposed with respect to the host galaxy or quasar, and, according to the standard picture, these are powered by beams or jets originating in an active galactic nucleus. The identification of flat-spectrum central radio cores in most of these sources has enabled a number of analyses of the kinematics of their lobes and hot-spots to be undertaken (Longair \& Riley 1979, Zieba \& Chyzy 1991, Best {\it et al.} 1995). Structural asymmetries and misalignments of the hot-spots and radio lobes have been used to investigate their intrinsic properties and to test unification schemes for radio quasars and radio galaxies (Scheuer 1987; Barthel 1987, 1989). 

In the simplest picture, it was conjectured that the radio lobes and hot-spots moved out symmetrically from an active nucleus at a significantly relativistic speed and the observed structural asymmetries could then be attributed to the differences in light travel times from the lobes to the observer (Ryle \& Longair 1967).  Many studies have been made of the probability distribution of the velocities of the radio source components from the observed distributions of the ratio of core--hot-spot distances (Longair \& Riley 1979; Katgert-Meikelijn {\it et al.} 1980; Banhatti 1980; Best {\it et al.} 1995).  The mean velocity of advance of the hot-spots was found to be $\geq 0.2c$, with a considerable spread about the mean velocity, some values greater than $0.4c$ being found.  

There are however significant problems with this model.  McCarthy {\it et al.} (1991) studied the spatial distribution of thermal emission-line gas about powerful FRII radio sources and found that the asymmetry in the distribution of the ionised gas was correlated with the structural asymmetry of the radio lobes, strongly suggesting that environmental asymmetries play a significant r\^ole.  This correlation between optical and radio asymmetries can be naturally attributed to clumpy environmental effects (Pedelty {\it et al.} 1989a, b).   A further possibility is that the jets of powerful FRII sources might well be intrinsically asymmetric. The  analysis of Wardle \& Aaron (1997) of the jet-counterjet flux ratios of 13 3CR quasars observed with high resolution by the VLA by Bridle {\it et al.} (1994) showed, however, that, where jets and counterjets are observed in quasars, the jet-counterjet brightness ratios can be attributed almost entirely to relativistic beaming and, at most, modest intrinsic asymmetries in the kiloparsec-scale jets are allowed. 

Further problems arise from observations of one-sided radio jets, which have now been observed in many FRII radio sources.  Following Wardle \& Aaron (1997), it is natural to attribute this one-sidedness to the effects of relativistic beaming.  Consequently, the lobe on the same side as the jet should be approaching the observer and should be longer than the lobe on the counterjet side.  Significant discrepances from this rule have been found for a number of FRII radio galaxies and quasars (Saikia 1981, 1984; Black, {\it et al.} 1992; Fernini, {\it et al.} 1993, 1997; Bridle {\it et al.} 1994; Scheuer 1995; Leahy, {\it et al.} 1997; Hardcastle {\it et al.} 1997, 1998). A further discrepancy is that the expansion speeds of the FRII radio sources found from the simple model are systematically greater, by at least a factor of two, than those found from spectral ageing arguments. This problem is discussed in Section 2.2. 

Best {\it et al.} (1995) realised that it is likely that both relativistic, environmental and intrinsic asymmetries play a r\^ole in determining the observed structures of FRII radio sources. To test the likely contributions of relativistic and environmental/intrinsic effects upon the observed structures of FRII sources, an \emph{asymmetric relativistic model} has been developed in which account is taken of the contribution of both relativistic and intrinsic/environmental asymmetry effects. By the term `relativistic' we mean that light-travel time differences play a significant r\^ole in determining the observed structural asymmetries.   The key parameter in the present analysis is the {\it jet-side} of the radio sources, which is estimated for $\sim 80\%$ of those in the 3CRR complete sample (Section 4).

In Section 2, the problems of the symmetric relativistic model are reviewed and, in Section 3, the asymmetric relativistic model is analysed quantitatively.  The sample of FRII radio sources from the 3CRR catalogue is described in Section 4. In Section 5, correlations of source asymmetry with radio luminosity and linear size are discussed.  A statistical analysis of the properties of FRII sources in the context of the asymmetric relativistic model is discussed in Section 6. The mean expansion speed of the lobes and the critical angle for orientation-based unification schemes for radio galaxies and quasars are the subjects of Section 7.  Except where otherwise stated, Hubble's constant has been taken to be $H_0=50$ km ${\rm s}^{-1}$ ${\rm Mpc}^{-1}$ and the deceleration parameter $q_0=0.5$.  

\section{Symmetric model}

In the symmetric relativistic model, the two hot-spots recede at a constant speed $v_0$ in opposite directions from an active galactic nucleus and any observed asymmetry is attributed to differences in light travel times to the observer. According to this model, the hot-spot approaching the observer at projected distance, $r_{\rm a}$, from the nucleus is older, and observed further from the nucleus, than the receding hot-spot which has projected distance $r_{\rm r}$.

Banhatti (1980) introduced the \emph{fractional separation difference}, $x=(r_{\rm a}-r_{\rm r})/(r_{\rm a}+r_{\rm r})$, to describe the structural asymmetry of the radio hot-spots. According to the model, $x$ is proportional to radial component of the velocity of hot-spots,

\begin{equation}
x \equiv \frac{r_{\rm a}-r_{\rm r}}{r_{\rm a}+r_{\rm r}}=\beta\cos{\theta},
\end{equation}
where $\beta=v_0/c$, $c$ is the speed of light and $\theta$ is the angle between the radio axis and the line of sight to the observer.

Assuming that the orientations of radio axes with respect to the line of sight are random, the simple integral equation relating the observed probability distribution of the radial velocities $x$, $g(x)$, to the probability distribution of true space velocities of the hot-spots $G(\beta)$ is
\begin{equation}
g(x)=\int_{x}^{1}\frac{G(\beta)}{\beta}\,{\rm d}\beta.
\end{equation} 
(Banhatti 1980).  The solution of this equation is
\begin{equation}
G(\beta)=-\,\beta\,g'(\beta).
\end{equation}
The determination of $G(\beta)$ was the subject of the papers by Banhatti (1980) and Best {\it et al.} (1995). 

\subsection{Mean and dispersion of hot-spot velocities}

It is useful to introduce moments of the probability distribution $G(\beta)$. Multiplying (3) by $\beta^n\,{\rm d}\beta$ and integrating from zero to one, the $k^{\rm th}$ moment of the observed fractional separation difference, 
\begin{eqnarray}
\nu_k=\frac{\nu_{0k}}{k+1},              
\end{eqnarray}
where the $k^{\rm th}$ moment of the distribution function of $\beta$ is
\begin{equation}
\nu_{0k}=\int_0^{1}\beta^k G(\beta)\,{\rm d}\beta.            
\end{equation}
The first and second moments of $G(\beta)$ and $g(x)$ are equal to the mean velocity and the mean square velocity, that is,
$\nu_{01}=\overline{\beta}$, $\nu_1=\overline{x}$, $\nu_{02}=\overline{\beta^2}$ and $\nu_2=\overline{x^2}$. Hence the mean space velocity and mean square velocity are,
\begin{equation}
\overline{\beta}=2\overline{x}\quad {\rm and}\quad\overline{\beta^2}=3\overline{x^2}
\end{equation}
The variance of the velocity about its mean value is 
\begin{equation}
\sigma_{\beta}^2=\overline{\beta^2}-\overline{\beta}^2=3\,\overline{x^2}-4\,\overline{x}^2.
\end{equation}

\subsection{Problems with the symmetric model}

\subsubsection{Expansion speed discrepancies}

\begin{figure}
\begin{center}
\epsfig{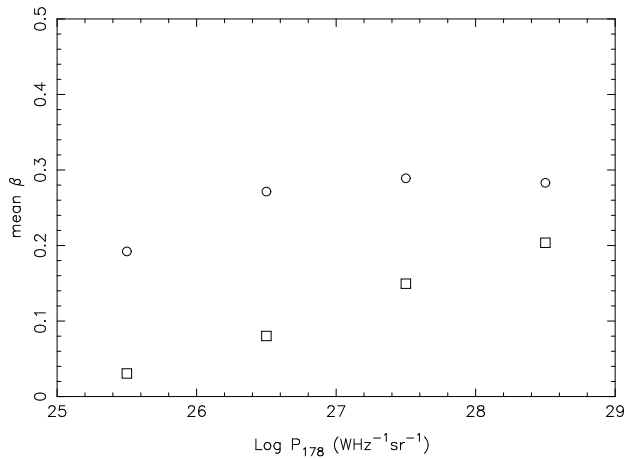}
\end{center}
\caption{The mean expansion speeds of the radio lobes $\overline{\beta}$ (in units of the speed of light) according to the symmetric relativistic model (circles) and from spectral ageing analyses (squares) plotted against radio power. The spectral ages were taken from the analyses of 33 3CRR sources of Alexander \& Leahy (1987) and Liu {\it et al.} (1992) and kinematic velocities from the symmetrical relativistic model for 132 sources in the complete 3CRR sample.} \label{fig-1}
\end{figure}  

The mean intrinsic speeds of the source components according to the symmetric relativistic model can be compared with those derived from synchrotron spectral ageing arguments. The latter speeds were estimated for 33 3CRR FRII radio sources in the power range $10^{25}\leq{P_{178}}\leq10^{29}\,{\rm W}{\rm Hz}^{-1}{\rm sr}^{-1}$ by Alexander \& Leahy (1987) and Liu {\it et al.} (1992), assuming that the source axes lay in the plane of sky, thus providing estimates of the tangential velocities ($y = v_{\rm t}/c$) of the lobes.  The true mean speed $\overline\beta$ is related to $\overline y$ by 
\begin{equation}
\overline{\beta}=\frac{4}{\pi}\,\overline{y}.
\end{equation}
In the same way, the standard deviations of the projected and true velocity distributions are related by 
\begin{equation}
\sigma_{\beta} = \sqrt{\,\frac{3}{2}\,\overline{y^2}-\left(\frac{4}{\pi}\,\overline{y}\right)^2},
\end{equation}
where $y=v_{\rm t}/c$.
The true mean speed of the source components derived from spectral ageing arguments from the above papers is $\overline{\beta}=0.13\pm0.08$, compared to a mean value of $\overline{\beta}=0.27\pm0.16$ from the observed distribution of fractional separation differences for the 132 3CRR FRII sources. Thus, the mean speed and its standard deviation of the source components are about two times greater than the same quantities derived from spectral ageing arguments. 

To check this result, the source sample described in Section 4 was divided into four equal logarithmic bins in radio power and mean velocity estimates were made using both techniques (Fig.~\ref{fig-1}).  In this analysis, the cosmologies adopted by the above authors have been adopted,  $H_0=50$ km ${\rm s}^{-1}$ ${\rm Mpc}^{-1}$ and deceleration parameter $q_0=0$. There were roughly equal numbers of sources in each of the bins.  For all four bins, the mean speeds estimated from the relativistic symmetric model are systematically greater, by factors of between 1.5 and 5, than the mean speeds obtained from the spectral ageing analyses. The differences in the estimates becomes smaller for the most powerful sources.  Whilst recognising that there are a number of reasons why the spectral ageing estimates might be in error, these discrepancies suggest that the velocities derived from the symmetric model are overestimated. 
 
The obvious interpretation of these data is that not all the source asymmetry can be attributed to the relativistic bulk motion of the source components.  On the other hand, the discrepancy is less than a factor of two for the most luminous sources and so, although the asymmetries may not be wholly attributed to relativistic effects, they cannot be neglected. 

\subsubsection{Jet-side discrepancies}

Laing (1988) and Garrington {\it et al.} (1988) discovered the important correlation between the degree of depolarisation of a radio lobe and the presence of a one-sided radio jet emanating from the active nucleus. The sense of the correlation is such that the jet-side lobe is significantly less depolarized than the counterjet-side lobe and can be attributed to the fact that the radiation from the approaching lobe passes through a smaller path length of the depolarising halo surrounding the radio source than does the receding lobe.  The Laing-Garrington effect thus enables the approaching and receding lobes to be identifed and suggests that the one-sidedness of the jet is the result of relativistic beaming.   The lobe approaching the observer should be longer than the receding lobe and should be the side on which the one-sided radio jet is found. 

Saikia (1981) used the jet-side to test the symmetric model, initially for a small sample of radio galaxies, and later for a larger sample of 36 quasars (Saikia, 1984). For about half the quasars in his sample, the jet was found to lie on the shorter side.  The same effect was also present in the larger sample of quasars and radio galaxies selected by Scheuer (1995).  The latest observations of jets from high-resolution VLA observations indicate that, in sources with double-sided jets, there is no tendency for the brighter to lie in the longer radio lobe.  According to the studies listed in Section 1, only 18 out of 33 radio galaxies, or 55\%, show the expected correlation and, in the quasar sample of Bridle {\it et al.} (1994), only 6 out of 13 sources follow the expectation of the relativistic symmetric model. 
       
\section{The asymmetric relativistic model}

\begin{figure}
\begin{center}
\epsfig{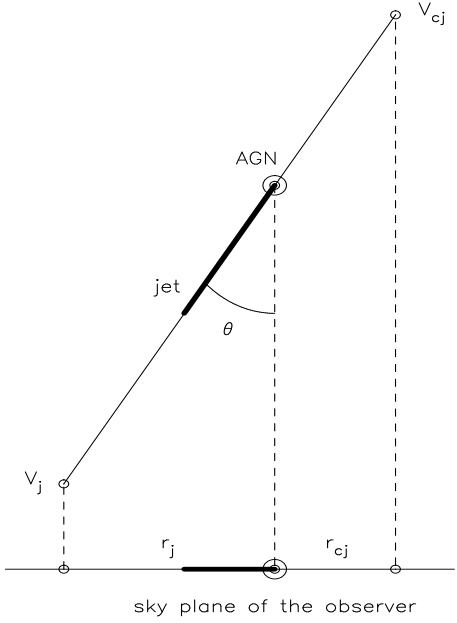}
\end{center}
\caption{The asymmetric relativistic model of FRII radio sources.} \label{fig-2}
\end{figure}

The \emph{asymmetric relativistic model} is designed to take account of the contributions of both relativistic and intrinsic/environmental asymmetries to the structures of the radio sources.  The jets advance through a clumpy asymmetric environment at an angle $\theta$ to the line of sight (Fig.~\ref{fig-2}).  Asymmetries associated with both the environment and the intrinsic properties of the jets can be described by assuming that the mean velocity of the lobe in the jet (approaching) direction $v_{\rm j}$ and that in the counterjet (receding) direction $v_{\rm cj}$ are different. Throughout this analysis, it is assumed that the jet-side, or the brighter of a two-sided jet, lies in the lobe approaching the observer and we will consistently use the term `jet-side' to have this meaning, even if the jets are absent, but the orientiation of the source can be found from other arguments. 

As in the symmetric case, a \emph{fractional separation difference} $x$ can be defined, but now the relationship between $x$, $v_{\rm j}$ and $v_{\rm cj}$ is slightly more complex than (1):
\begin{equation}
x \equiv \frac{r_{\rm j}-r_{\rm cj}}{r_{\rm j}+r_{\rm cj}} = \frac{v_{\rm j}-v_{\rm cj}}{v_{\rm j}+v_{\rm cj}}+\frac{2}{c}\,\frac{v_{\rm j}\,v_{\rm cj}}{v_{\rm j}+v_{\rm cj}}\,\cos\theta,
\end{equation}
where $r_{\rm j}$ and $r_{\rm cj}$ are the projected distances between the active nucleus and the furthest ends of the lobes on the jet and counterjet sides respectively, and $c$ is the speed of light.

In the symmetric relativistic model, $v_{\rm j} = v_{\rm cj} = v_0$ and the value of $x$ is always positive, $x \in [0;1]$. In the asymmetric relativistic model, however, negative values of $x$ can be found, $x \in [-1;1]$. Specifically,  if the first term on the right-hand side of (10) is negative and of greater magnitude than the second, negative values of the fractional separation difference are found.

The speeds $v_{\rm j}$ and $v_{\rm cj}$ can be written, $v_{\rm j}=v_0 + v_{\rm jd}$ and $v_{\rm cj}=v_0 + v_{\rm cjd}$, where $v_{\rm jd}$ and $v_{\rm cjd}$ are the changes to $v_0$ due to intrinsic/environmental asymmetries on the jet and counterjet sides respectively. Introducing new variables $\delta_{\rm j}=1+v_{\rm jd}/v_0$ and $\delta_{\rm cj}=1+v_{\rm cjd}/v_0$, (10) can be written
\begin{equation}
x = \frac{\delta_{\rm j}-\delta_{\rm cj}}{\delta_{\rm j}+\delta_{\rm cj}}+\frac{2}{c}\,\frac{\delta_{\rm j}\,\delta_{\rm cj}}{\delta_{\rm j}+\delta_{\rm cj}}\,v_0\,\cos\theta.
\end{equation}  
In the symmetric case, $\delta_{\rm j} = \delta_{\rm cj} = 1$, we recover (1) and $v_0$ is the average speed of advance of the lobes. In the asymmetric model, the same interpretation is correct, provided the mean value of the joint distribution of $\delta_{\rm j}$ and $\delta_{\rm cj}$ is unity.  In other words, $v_0$ is the average advance speed of the lobes, provided averages are taken over large samples of sources normalised so that $\langle\delta\rangle = 1$. In physical terms, we would expect $v_0$ to depend on the mean jet-counterjet power and the mean environmental density, whereas $\delta_{\rm j}$ and $\delta_{\rm cj}$ are measures of the intrinsic asymmetry of the jets and inhomogeneities in the environments of the radio sources.

Let us examine (11) for positive and negative values of $x$.
\begin{itemize}
\item If $\delta_{\rm j}-\delta_{\rm cj}\geq0$, $x\geq0$ for all $\theta$. 
\item If $\delta_{\rm j}-\delta_{\rm cj}\leq0$, two cases are possible: $x\geq0$ for $\theta < \theta_{\rm n}$ and $x\leq0$ for $\theta > \theta_{\rm n}$ where
\begin{equation}
\theta_{n}=\arccos\left(\frac{c}{v_0}\frac{|\delta_{\rm j}-\delta_{\rm c  j }|}{2\,\delta_{\rm j}\delta_{\rm cj}}\right).
\end{equation}
\end{itemize}

\begin{figure}
\begin{center}
\epsfig{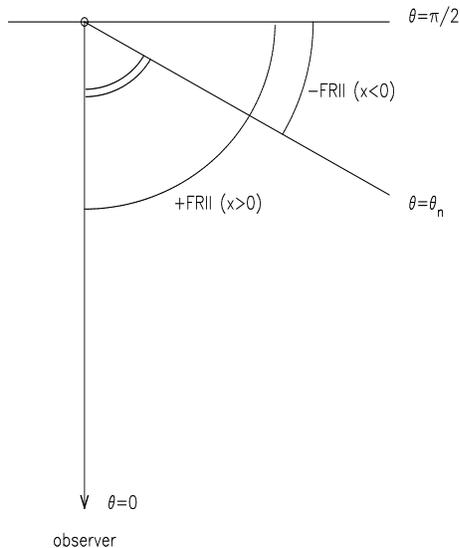}
\end{center}
\caption{The orientation scheme for FRII radio sources.}\label{fig-3}
\end{figure}

The viewing angles for FRII sources with the jet on the long lobe side ($x\geq0$, hereafter, +FRII sources) span all angles from $0$ to $\pi/2$, whereas the viewing angles for FRII sources with the jet on the short lobe side ($x\leq0$, hereafter $-$FRII sources) are restricted to the range $\theta_{\rm n}\leq\theta\leq\pi/2$ (see Fig.~\ref{fig-3}). 
If relativistic effects are negligible ($v_0\ll{c}$), the value of $\theta_{\rm n}\sim 0$; if relativistic effects are the predominant cause of the observed asymmetries, $\theta_{\rm n}\rightarrow{\pi/2}$. Thus, the radio axes of \emph{relativistic} $-$FRII sources lie preferentially close to the plane of sky, while the axes of \emph{non-relativistic} $-$FRII sources can be seen at almost all angles. Notice that, if the sources are non-relativistic, $v_0 \ll c$, and are significantly asymmetric intrinsically, the argument of arccos in (12) can become greater than 1 and then negative values of $x$ can be found for all angles in the range $0 < \theta < \pi/2$. For sources in which the relativistic effects are dominant, these should be greater for the +FRII sources than for $-$FRII sources; the intrinsic/environmental asymmetry is more significant for $-$FRII sources and should be dominant at low velocities for both +FRII and $-$FRII sources.  Large positive values of $x$ are found in FRII sources which expand with the highest speeds $v_0$ at angles close to the line of sight $\theta\sim 0$, while large negative values of $x$ are associated with intrinsically asymmetric sources lying close to the plane of the sky. 

Let us consider how the ratio of the numbers of $-$FRII to +FRII sources depends upon the value of $v_0$.  As $v_0\rightarrow 0$, $x$ is an indicator of the intrinsic/environmental asymmetry of FRII sources, $x \sim (v_{\rm jd}-v_{\rm cjd})/(v_{\rm jd}+v_{\rm cjd})$. If the jet axes and the intrinsic/environmental asymmetries are distributed isotropically, the velocities  $v_{\rm jd}$ and $v_{\rm cjd}$ on the jet and counterjet sides should be randomly distributed with the same probability distribution. Then, the ratio of the number of $-$FRII to +FRII sources should be equal, that is,  $N(-\mbox{FRII})/N(+\mbox{FRII})\sim 1$.  If the relativistic effect is dominant, $v_{\rm jd}$, $v_{\rm cjd} \ll v_0$ and $x\rightarrow \cos\theta\geq 0$. In this case, the ratio $N(-\mbox{FRII})/N(+\mbox{FRII})$ tends to $0$. We can define an \emph{asymmetry parameter} $\varepsilon$, where
\begin{equation}
\varepsilon=1-2\,\frac{N(-\mbox{FRII})}{N(+\mbox{FRII)}}.
\end{equation}         
Relativistic effects are more important than intrinsic/environmental asymmetries if $0<\varepsilon\leq1$, while, if $-1\leq\varepsilon<0$, the latter effects are dominant. If $\varepsilon \sim 0$, the contribution of each effect is of comparable importance.     

Let us consider how the observed distribution of $x_{\rm l}$ depends upon the properties of the asymmetric model.  The mean speeds of the jet and counterjet lobes are the sums of the mean intrinsic speed of the lobes, $v_0$, and the random dispersion about the mean, $v_{\rm j} = v_0 + v_{\rm jd}$ and $v_{\rm cj} = v_0 + v_{\rm cjd}$.  If we assume that  the intrinsic asymmetry is distributed isotropically in the sky, the distribution function of the space velocities of the lobes on the jet- and counterjet-sides should be the same, $F(v_{\rm j}) = F(v_{\rm cj}) = F(v_{\rm l})$, implying that the distribution functions $G(v_{\rm jd}) = G(v_{\rm cjd}) = G(v_{\rm d})$.  

For illustrative purposes, let us assume that (i) the distribution function of the intrinsic velocities $v_0$ is Gaussian with the mean $\overline{v}_0$ and standard deviation $\sigma_{v_0}$ and (ii) the distribution function of random disturbed velocities is also a Gaussian with constant standard deviation $\sigma_{v_{\rm d}}$ on the jet- and counterjet sides.  We assume that these normal distribution functions are independent. Then, the distribution function of their sum, the lobe speeds $v_{\rm j}$ and $v_{\rm cj}$ are also normal with a mean speed $\overline{v}_0$ and dispersion $\sigma_{v_{\rm l}}^2=\sigma_{v_0}^2+\sigma_{v_{\rm d}}^2$,
\begin{equation}
F(v_{\rm l})=F(v_0+v_{\rm d})={\frac{1}{\sigma_{v_{\rm l}}\sqrt{2\pi}}\,\exp^{-\frac{(v_{\rm l}-\overline{v}_0)^2}{2\,\sigma_{v_{\rm l}}^2}} }.
\end{equation}
For illustrative purposes, let us adopt a mean lobe speed $\overline{v}_{\rm l} = 0.15 c$ and $\sigma_{v_{\rm l}} = 0.05c$. Then, in Fig.~4 and Table 1, the predicted distributions of $x_{\rm l}$, and the mean values of $\varepsilon$ and $x_{\rm l}$, are shown for different combinations of the intrinsic dispersions in the lobe speeds $\sigma_{v_0}$ and the intrinsic/environmental asymmetry $\sigma_{v_{\rm d}}$.   The first and fourth lines of Table 1 correspond to the limiting cases of $\sigma_{v_{\rm d}} \approx 0$ and 0.05 respectively.

\begin{figure}
\begin{center}
\epsfig{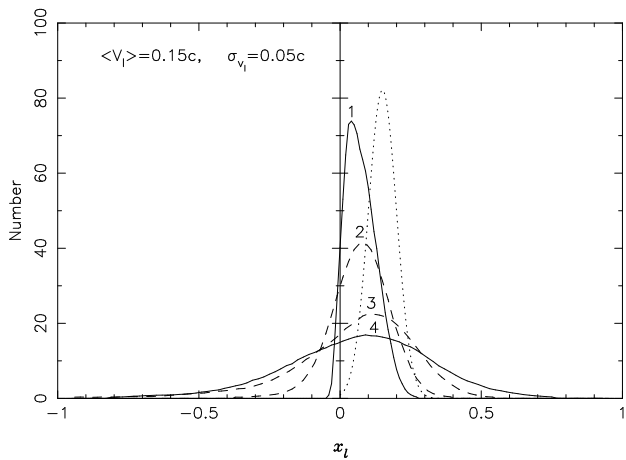}
\end{center}
\caption{The effects of changing the $\sigma_{v_{\rm d}}$ upon the observed distribution of $x_{\rm l}$.  The dotted line shows the distribution of lobe velocities with  mean $\overline{v}_{\rm l} = 0.15c$ and standard deviation $\sigma_{v_{\rm l}} = 0.05c$. The lines labelled 1, 2, 3 and 4 show the predicted distributions of $x_{\rm l}$ for different combinations of $\sigma_{v_0}$ and $\sigma_{\rm v_{\rm d}}$ and are listed in Table 1.}\label{fig-4}
\end{figure}

It can be seen that the distribution function of $x_{\rm l}$ becomes more asymmetric for large values of $\sigma_{v_{\rm d}}$, leading to a decrease in the value of $\overline{x}_{\rm l}$.  Therefore the simple relation (6) between the mean speed and the mean of the observed fractional separation difference underestimates $\overline{v}_{\rm l}$ and should be written 
\begin{equation}
\overline{v}_{\rm l}/c = \overline{\beta} \ga 2\overline{x}_{\rm l}.
\end{equation}
The asymmetry parameter is a better indicator of the intrinsic/environmental asymmetry than $\overline{x}_{\rm l}$ (see Table 1). Notice that the value $\varepsilon \approx 0$ is attained when $\sigma_{v_{\rm d}} \approx \sigma_{v_0}$, that is, the intrinsic/environmental and relativistic effects are of comparable importance. 
\begin{table}
\caption{Illustrating the effect of varying the magnitude of the intrinsic/environmental asymmetry for fixed values of $\overline{v}_{\rm l} = 0.15 c$ and $\sigma_{v_{\rm l}} = 0.05c$} 
\begin{center} 
\begin{tabular}{ccccc}  
Model& $\sigma_{v_0} $ & $\sigma_{v_{\rm d}}$ & $\varepsilon$ &  $\overline{x}_{\rm l}$ \\ \hline \hline
1&0.049& 0.001  &0.95&0.075         \\  
2&0.048 & 0.015 &0.48 & 0.075            \\ 
3&0.035&0.035 & 0 & 0.072\\
4&0.001&0.049&-0.18&0.071              
\end{tabular}
\end{center}
\end{table}

It should be noted that the asymmetry parameter $\varepsilon$ is a rather crude quantitative measure of the role of intrinsic/environmental asymmetries because of the small number statistics generally involved in determining $N(-{\rm FRII})$ and $N(+{\rm FRII})$, resulting in quite large error bars in the estimates of $\varepsilon$. 

\section{Sample of FRII radio sources}

\subsection{Statistics}

The analysis was based upon an analysis of the sample of 132 FRII 3CR radio sources selected from the catalogue of Laing, Riley and Longair (1993) which lay in the area of sky $\delta \ge 10^{\circ}$ and $|b| \ge 10^{\circ}$. The jet-side could be determined for 103 of the 3CRR sources using the following criteria in order of preference:  
\begin{itemize}
\item {\it Direct VLA/VLBI images of kiloparsec and/or parsec-scale jets} (82\%). As discussed by Hardcastle {\it et al.} (1998), where high resolution, high sensitivity maps of sources are available, well-defined radio jets are remarkably common. We use the term `definite' jets (D) when a jet can be defined according to the definition of Bridle and Perley (1984), namely, that the radio feature is at least four times longer than it is wide.  In a number of cases, evidence for radio jets has been found at a reasonable level of confidence within the extended radio source structure and we refer to these as `probable' jets: these correspond to the `possible' jets of Hardcastle {\it et al.} (1997) and the `candidate' jets of Fernini {\it et al.} (1997).  Furthermore, Pearson \& Readhead (1988) found that, in cases where both a VLBI (parsec-scale) jet and kiloparsec jet are seen, the VLBI jet is connected with, or is on the same side as, the kiloparsec jet. The same result was found in the sample of sources studied by Wardle \& Aaron (1997). Parsec-scale jets could therefore be used to define the jet-side.
\item {\it Depolarization asymmetry -- the Laing-Garrington effect}  (13\%). It is assumed that the less depolarized lobe is approaching the observer. 
\item {\it Spectral index asymmetry -- the Liu-Pooley effect} (5\%). Liu \& Pooley (1991) found a strong correlation between the depolarization and spectral index asymmetries of the lobes for a small sample of FRII radio sources, in the sense that the lobe which is less depolarized has the flatter spectrum. The lobe with the flatter spectrum is assumed to be approaching the observer.
\end{itemize}
The remaining 29 sources could not be classified for a number of reasons. In a number of cases, excellent radio maps were obtained by Fernini {\it et al.} (1993, 1997), but no evidence for jets could be discerned.  In other cases, radio maps of sufficiently high quality have not been published. 

Scheuer (1995) used the statistics of the structural asymmetries of the {\it radio lobes}, rather than {\it hot-spots}, to define the fractional separation differences, arguing that the hot-spots are ephemeral structures which can change their location within the volume of the lobe; the largest lobe length is therefore likely to be a more stable measure of the mean growth rate of the lobe.  Means and standard deviations of the fractional separation differences for the lobes and hot-spots in the present sample are $\overline{x}_{\rm l}=0.07 \pm 0.18$ and $\overline{x}_{\rm h}= 0.06 \pm 0.22$, respectively. The greater standard deviation of ${x}_{\rm h}$ as compared with that of ${x}_{\rm l}$ shows that the locations of hot-spots are on the average more asymmetric than the lobe lengths, consistent with Scheuer's argument.  For this reason, the fractional separation difference of lobes has been used in the rest of the analysis.   It should be noted that the larger dispersion in $x_{\rm h}$ as compared with $x_{\rm l}$ means that the high mean velocities of expansion found by Longair \& Riley (1979), Banhatti (1980) and Best {\it et al.} (1995) are overestimates.  Details of the sources included in this analysis and the criteria used, as well as the adopted values of the fractional separation differences $x$ for the lobes and hot-spots are given in the Appendix.
  
Of the 103 FRII sources, 71 were radio galaxies and 32 were quasars. Among the 71 {\it radio galaxies}, the jet-side was determined (i) by the presence of jets (34 cases) and probable jets (21 cases), (ii) from depolarization asymmetries (12 cases) and (iii) from spectral index asymmetries (4 cases).   Among the 32 {\it quasars}, 28 cases of jets and two probable jets were found; the jet-side was found from depolarization asymmetries and from spectral index asymmetries in one case each.  Two jets among the 36 radio galaxies were detected on the parsec-scale from VLBI images. 

Among the 103 sources, there are 67 +FRII sources (65\%) and 36 $-$FRII sources (35\%). Of the 71 radio galaxies, 43 are +FRII and 28 are $-$FRII sources; among the 32 quasars, 24 are +FRII and 8 $-$FRII sources.

\begin{figure}
\begin{center}
\epsfig{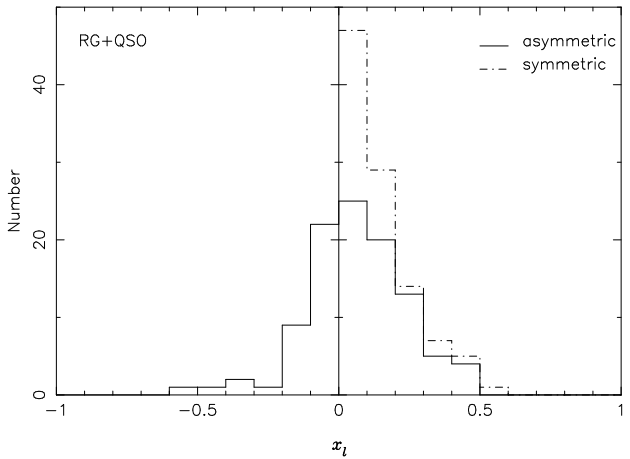}
\end{center}
\caption{The distribution of fractional separation differences according to the asymmetric (full) and symmetric (dot-dash) models for 103 FRII 3CR radio sources.}\label{fig-5}
\end{figure}  

Fig.~\ref{fig-5} shows the distributions of the fractional separation differences for the symmetric ($0<|x_{\rm l}|<1$) and asymmetric pictures ($-1<x_{\rm l}<1$).  The latter distribution, containing 36 $-$FRII sources (35\%), is asymmetric, suggesting that both intrinsic/environmental and relativistic effects are important.

\subsection{Selection effects}

The 3CRR sample is a complete unbiased sample of sources selected at a low radio frequency. It is orientation-independent since the flux densities of the sources in the sample are generally dominated by the diffuse emission of the radio lobes. The presence of jets or probable jets has been found in virtually all 32 quasars in the sample and in 70\% of the radio galaxies. Therefore, the uncertainties are associated with the presence or otherwise of jets in the remaining 20\% of the complete sample and these are not likely to be large. 

Let us consider first the reliability of using the `probable' jets in the analysis.  The sample was divided into two subsamples, the first containing all sources with probable jets, and the second the remaining 84 sources with definite jets.  The statistics of probable jets shows that 16/22 (72\%) are +FRII sources while 6 (28\%) are $-$FRII sources. For the second sample, 51/81 (62\%) and 30/81 (38\%) are +FRII and $-$FRII sources respectively.  We conclude that no bias is introduced by including the probable jets in the analysis. 
 
According to the relativistic beaming model for the radio jets, there is a strong selection effect favouring the detection of jets in FRII sources which are oriented close to the line of sight. Within the context of orientation-based unified schemes, it is therefore natural that the sources without detectable jets should be radio galaxies in which the axis of ejection lies close to the plane of the sky. For these `missing' sources, the fractional separation difference should be predominantly due to intrinsic/environmental asymmetries and the distributions of the space velocities of lobes $v_{\rm j}$ and $v_{\rm cj}$ should be symmetric about $x_{\rm l} = 0$. Therefore, jet orientation selection effect would not change the symmetry of the observed distributions of $x_{\rm l}$, but would slightly decrease the asymmetry parameter.  In the worst case, the 29 sources for which the jet side has not been determined would be distributed exactly symmetrically and so we would expect 14.5 more $-$FRII and 14.5 more $+$FRII sources, reducing the estimate of $\varepsilon$ from the value $\varepsilon = -0.3$ found in Section 6.1 to $-0.47$.

The sample of 103 sources was derived from a literature search of all available radio maps and it is important to compare the statistics with samples which have been observed with more uniform criteria of sensitivity and angular resolution.  High-resolution observations of a sample of 50 3CR FRII radio galaxies with $z < 0.3$ are presented by Hardcastle {\it et al.} (1998), representing the combined samples of Black {\it et al.} (1992), Leahy {\it et al.} (1997) and Hardcastle {\it et al.} (1997).  34 of the 50 FRII radio galaxies belong to the 3CRR complete sample. Of these 34 radio galaxies, only 6 (or 18\%) did not possess jets or probable jets, a somewhat greater success rate than for our sample, as expected.     

For 12 radio galaxies, depolarisation asymmetries have been used to determine the jet-side where no radio jet has been detected.  Since no jet was detected, the axes of these sources are likely to lie close to the plane of the sky and intrinsic/environmental asymmetries are likely to be important.  These may also be more important than the Laing-Garrington effect in causing the observed depolarisations (Pedelty {\it et al.} 1989a and McCarthy {\it et al.} 1991). Taking the jet-side to be the less depolarized side may change the true sign of $x_{\rm l}$ from positive to negative or vice versa with equal probability, as the orientation of the jet axes are assumed to be isotropic. This selection effect is not likely to influence strongly the observed distribution of $x_{\rm l}$ and the asymmetry parameter because it concerns only 12 sources.  We have repeated the complete analysis excluding these 12 radio galaxies and found that their exclusion makes no difference to our conclusions. 
  
Scheuer (1995) noted a number of other effects which could bias the analysis of the jet-side of the FRII sources. For example, misalignment of the jets, intrinsic one-sidedness of the radio jets, and misalignment and distortion of the radio lobes, can all complicate the analysis.  These could all be modelled (see for example Best {\it et al.} 1995), but the effects were found to be quite small in that analysis. 

\section{Fractional separation differences of the radio lobes}

\subsection{Correlations with radio luminosity}

\begin{figure*}
\begin{center}
\vbox to100mm{\vfil
\epsfig{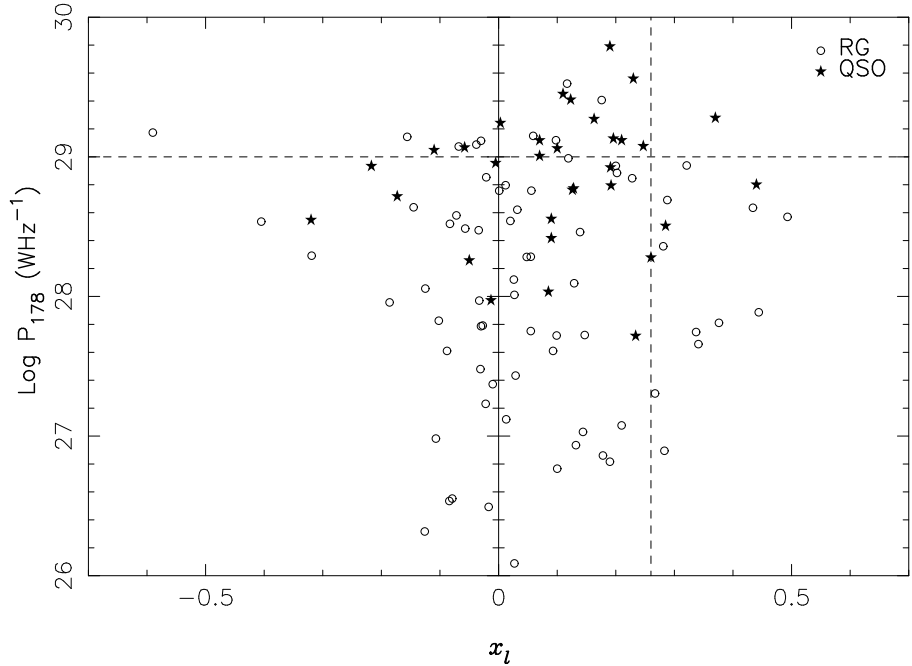}
\caption{The logarithm of radio luminosity plotted against fractional separation difference for 71 radio galaxies and 32 quasars.}\label{fig-6}
\vfil}
\label{landfig}
\end{center}
\end{figure*} 

\begin{figure*}
\begin{center}
\vbox to100mm{\vfil
\epsfig{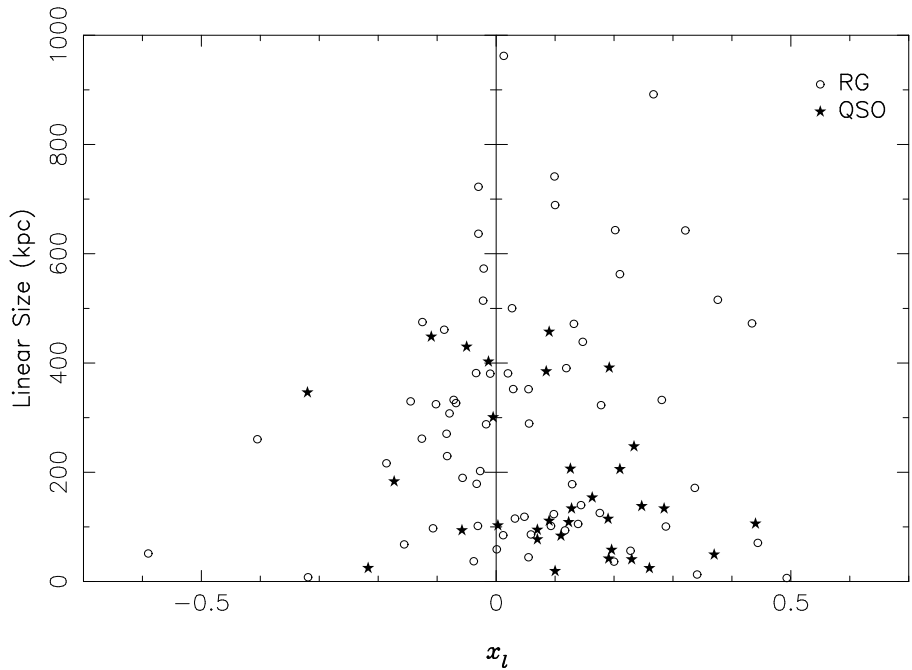}
\caption{The linear size against fractional separation difference $x_{\rm l}$ for the same sample as in Fig.~5.}\label{fig-7}
\vfil}
\label{landfig}
\end{center}
\end{figure*}  

A plot of radio luminosity against fractional separation difference is presented in Fig.~\ref{fig-6}. The diagram is V-shaped with a tendency for the spread in fractional separation difference to increase with increasing luminosity up to $P_{178} \approx 10^{29}{\rm W} {\rm Hz}^{-1}$. Intrinsic/environmental asymmetries are necessary to account for negative values of $x_{\rm l}$ and there is a trend for this asymmetry to increase with luminosity, or with redshift since, in the present flux density limited sample, radio luminosity and redshift are correlated. A possible explanation for this increase is that high redshift ($z>0.5$) FRII radio galaxies are associated with the brightest galaxies in rich clusters, while their low redshift counterparts tend to lie in isolated environments or in small groups (Lilly \& Prestage 1987, Hill \& Lilly 1991, Best {\it et al.} 1998). Gradients in the intergalactic gas density should be present in these rich clusters on the scale of the overall structure of the radio sources, which might lead to asymmetric environments on large scales. Furthermore, Best {\it et al.} (1996, 1997) present evidence that the interstellar medium in the vicinity of 3CR radio galaxies at $z \sim 1$ must contain clumpy cold gas to account for the pronounced alignment effect displayed by these galaxies, which is not observed at small redshifts.  Both intrinsic/environmental and relativistic effects contribute to positive values of $x_{\rm l}$, and there is considerable spread of these values at high luminosities.

The luminosity--fractional separation difference correlation for high luminosity quasars $P_{178} > 10^{29}\,{\rm W}{\rm Hz}^{-1}$ is of special interest since jet-sides have been determined for all of them. There are two important features of their distribution. First, there are only 2 $-$FRII quasars and, second, there is only one +FRII quasars with $x_{\rm l} \ga 0.25$ (Fig.~\ref{fig-6}).   The lack of $-$FRII quasars suggests that the relativistic effect is dominant for luminous quasars, that the axes of quasars must lie at reasonably small angles to the line of sight and that the ratios $v_{\rm jd}/v_0$ and $v_{\rm cjd}/v_0$ should tend to zero.  The lack of quasars with large positive values of $x$ suggests that most values of $x$ must lie in the restricted range $0<x_{\rm l} \la {x_{\rm max}}$. In the case of the symmetric relativistic model, $x_{\rm l} = v_0\,\cos\theta/c$.   Recognising that there must be some residual environmental/intrinsic effect present at these high luminosities, a reasonable upper limit to the value of $x$ is $x_{\rm max}= 0.26$, which is shown by the vertical dashed line.  Then, assuming $\theta \sim 0$, we obtain an firm \emph{upper limit} to the expansion speed of the lobes of $v_0 \la{x_{\rm l}\,c} = 0.26c$ for high luminosity quasars.

At lower luminosities, the $+$FRII sources with $x_{\rm l} \ga 0.26$ and the $-$FRII sources may be accounted for by a combination of intrinsic/environmental asymmetries in mildly relativistic sources.  This picture is supported by the corresponding values of the asymmetry parameter $\varepsilon$.   For all quasars with $P_{178} > 10^{29}$ W Hz$^{-1}$, the asymmetry parameter is $\overline{\varepsilon}_{\rm Q}(>10^{29})\approx 0.7 \pm 0.24$, whereas for quasars with smaller luminosities, $\overline{\varepsilon}_{\rm Q}(<10^{29}) \approx - 0.1 \pm 0.5$.  Thus, the relativistic effects appear to be dominant at high luminosities, whereas at lower luminosities, both effects are of comparable importance.

This last result appears to differ from that found by Scheuer (1995) who found that relativistic effects are small for the most luminous sources. His analysis was based upon the ratios of lobe lengths for three samples of quasars, including a sample of 19 quasars from the LRL sample. For this sample, Scheuer found ratios of the arm lengths which correspond to $\overline{x}_{\rm l} = 0.097$, in excellent agreement with our estimate for all the quasars in our sample of $\overline{x}_{\rm l}({\rm QSO}) = 0.102$.  The value of the asymmetry parameter for the LRL quasars studied by Scheuer is $\varepsilon = 0.29$, compared with our value of 0.33.  Thus, there is no discrepancy between Scheuer's results and the present analysis for corresponding sets of data.  It may be significant that one of the other samples of discussed by Scheuer, the BMSL sample, was restricted to quasars with redshifts $z > 1.5$, whereas the present sample of 103 sources has only four objects with redshifts greater than this value.  We note that Scheuer's sample was largely restricted to quasars, whereas our sample includes all radio galaxies and quasars in the 3CRR sample and should be free from systematic bias. We also note that the results of multi-epoch global VLBI observations of Compact Symmetric Objects, which are presumed to be young radio-loud sources, show strong evidence for mildly relativistic hotspot speeds, $v_{\rm h} > 0.1c$ (Owsianik \& Conway 1998, Owsianik {\it et al.} 1998). 

\subsection{Correlation with projected linear size}

Fig.~\ref{fig-7} shows a plot of projected linear size against fractional separation difference. This diagram has roughly the form of an inverted V-shape, which is not particularly surprising in view of the well-known inverse correlation between project linear size and radio luminosity.  The spread of negative values of $x_{\rm l}$ for small sources indicates that the intrinsic asymmetries are important on small physical scales, and many of these are associated with radio galaxies.   The majority of the quasars have positive values of $x$. There is some suggestion that the dispersion in $x$ becomes smaller at larger physical sizes.   The sample has been divided into two equal samples of small and large sources and these sub-samples have asymmetry parameters $\varepsilon_{\rm G+Q}(<250\, \mbox{kpc})\sim 0.3 \pm 0.2$ and $\varepsilon_{\rm G+Q}(>250\, \mbox{kpc})\sim-0.7 \pm 0.4$  respectively. These results indicate that relativistic and environmental/intrinsic effects are important for smaller physical sizes, while intrinsic symmetries are more important for larger sources. These differences in asymmetry parameters can be at least partly explained by projection effects.  If the larger sources lie predominantly close to the plane of the sky, they will be observed to be larger than those seen at a smaller angles to the line of sight and their structures would be dominated by intrinsic/environmental asymmetries.  The smaller sources are observed closer to the line of sight and so the relativistic effects are of greater importance, although the intrinsic asymmetries still play a significant r\^ole.  The fact that the quasars all lie towards the lower part of the diagram is consistent with an orientation-based unification picture.  Thus the results of Section 5.2 and 5.3 are not independent, but  are part of a consistent picture.    

\section{Statistical analysis}
 
\subsection{Asymmetry parameter}

The asymmetry parameter for the sample of 103 FRII sources is $\varepsilon_{\rm G+Q}=-0.07 \pm 0.22$, while for radio galaxies and quasars separately the values are $\varepsilon_{\rm G}=-0.3 \pm 0.32$ and $\varepsilon_{\rm Q}=0.33 \pm 0.36$ respectively. There are no quasars of low radio luminosity in the 3CRR sample and so a smaller sample limited to high luminosity sources $P_{178} \geq 10^{27.7}{\rm W}{\rm Hz}^{-1}$ has also been studied, this luminosity corresponding to the lower luminosity limit for quasars in the 3CRR sample.  The values of the asymmetry parameters for these samples are $\varepsilon_{\rm G+Q}=0 \pm 0.24$, $\varepsilon_{\rm G}=-0.27 \pm 0.36$ and $\varepsilon_{\rm Q}=0.33 \pm 0.36$ for 81, 49 and 32 sources respectively. Thus, the results for the whole and restricted samples are similar. These results suggest that intrinsic/environmental asymmetries and relativistic effects are of comparable importance for the sample overall, but the intrinsic asymmetries are more significant for the radio galaxies and relativistic effects are for the quasars, $\varepsilon_{\rm G}<\varepsilon_{\rm Q}$, consistent with the asymmetric relativistic model and in qualitative agreement with the expectations of orientation-based unification schemes.  Furthermore, the mean value of $x_{\rm l}$ for the sample of radio galaxies and quasars is $\overline{x}_{\rm G + Q} = 0.07 \pm 0.018$, while the values for radio galaxies and quasars separately are $\overline{x}_{\rm G} = 0.056 \pm 0.022$ and $\overline{x}_{\rm Q} = 0.102 \pm 0.028$, again consistent with the above picture.

\subsection{One- and two-sided jets}

For the sources in which jets have been detected, the proportion of two-sided jets is roughly the same for $-$FRII and +FRII radio galaxies -- 6/28, or 21\%, for $-$FRII and 8/43, or 19\%, for +FRII radio galaxies.   For the quasars, on the other hand,  4/8 (50\%) of the $-$FRII sources and 3/24 (13\%) of the +FRII sources  have two-sided jets. This difference between the fractions of $-$FRII and +FRII quasars is consistent with the $-$FRII quasars lying closer to the plane of the sky than the +FRII quasars, as well as with the expectations of the asymmetric relativistic model and with orientation-based unification schemes for radio galaxies and quasars. Of the 12 highest luminosity FRII quasars, 11 have one-sided jets and 1 is two sided (92\% and 8\%). The asymmetry parameter for these sources is $\varepsilon_{\rm Q}(P_{178}>10^{29})\approx 0.7\pm 0.2$ which strongly supports the supposition that relativistic effects are dominant for these sources.  The only anomaly is that the percentage of $-$FRII quasars with two-sided jets is greater than that of $-$FRII radio galaxies ($50\%>21\%$), since we would expect more two-sided jets among the $-$FRII galaxies. The small statistics of $-$FRII quasars may partly be the cause of this discrepancy. Furthermore, the detection of jets is strongly dependent upon the quality of the radio maps and generally the quasars have been studied with higher resolution and sensitivity than the radio galaxies.

\subsection{Mean linear size}

The mean linear sizes of $-$FRII and +FRII sources have been calculated for two samples.  The first consists of 19 low luminosity sources with $P_{178}<10^{27.7}\,{\rm W} {\rm Hz}^{-1}$, the lower radio luminosity limit for the quasars in the sample.  Three sources DA240, 3C236 and 3C326 were excluded from the calculation because of their extremely large linear sizes (2070, 5950 and 2660 kpc). For these low luminosity radio galaxies, the mean linear size of $(450 \pm 330)$ kpc for +FRII sources is larger than that $(300\pm140)$ kpc for $-$FRII sources (Table 2), where the `errors' are the standard deviations of the distributions. The large negative asymmetry parameter, $\varepsilon \sim -0.8$, for this low luminosity sample as a whole shows that intrinsic asymmetries play the dominant r\^ole in their structural asymmetries.  The inference is that the lobes of these radio galaxies have non-relativistic expansion speeds.  Furthermore, these low luminosity radio galaxies have hot-spot advance speeds determined from synchrotron ageing arguments which are significantly smaller than those of the higher luminosity sources, as can be seen from Fig.~1.     

For the sample of 81 high luminosity radio galaxies and quasars, the mean linear size of the $-$FRII radio galaxies is approximately equal to the mean linear size of +FRII radio galaxies $(290 \sim 280\, {\rm kpc})$, while for quasars the mean size of $-$FRII quasars is twice as large as the mean size of the +FRII quasars (Table 2). This is consistent with a model in which the axes of the radio galaxies lie at larger angles to the line of sight than those of the quasars, as illustrated in Fig.~8. For large angles, the projected size varies slowly and therefore the mean sizes of both $-$FRII and +FRII radio galaxies should be comparable on average. On the other hand, at smaller angles to the line of sight, the projected size varies more rapidly. According to the model, the  $-$FRII quasars are observed on average at greater angles ($\theta_{\rm c} > \theta > \theta_{\rm n}$),  than the +FRII quasars, ($\theta <\theta_{\rm n}$), and so should have larger projected sizes. 

\begin{figure}
\begin{center}
\epsfig{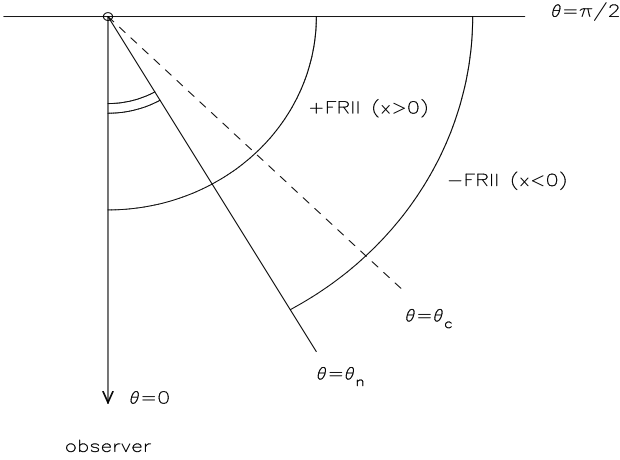}
\end{center}
\caption{Orientation of FRII radio sources according to the orientation-based unified scheme.  The value of $\theta_{\rm n}$ depends upon the randomly selected values of $v_0$, $v_{\rm jd}$ and $v_{\rm cjd}$. In orientation-based unified schemes, quasars are observed when the line of sight to the quasar lies in the angular range $0 \le \theta \le \theta_{\rm c}$ and radio galaxies when it lies in the range $\theta_{\rm c} \le \theta \le \pi/2$.}\label{fig-8}
\end{figure}  

The high luminosity sample has been divided into two equal high- and low-luminosity samples to test these predictions (Table 3). For high-luminosity radio galaxies, the difference between mean sizes of $-$FRII and +FRII sources ($280 > 220$ kpc) is greater than for low luminosities ($330 > 300$ kpc), suggesting that the difference in the mean sizes increases with luminosity. In addition, the asymmetry parameters are different for the two samples, $-0.44$ for the low luminosity sample and $\sim 0$ for the high luminosity sample, although the statistical significance is not great. This further supports the inference that relativistic effects are more important for the highest luminosity sources; this relation is consistent with the results of spectral ageing analyses which indicate that expansion speeds of lobes increases with luminosity (Fig.~1). 

\begin{table}
\caption{The mean projected linear sizes of the sources in the sample. The first line refers to radio galaxies with $P_{178} < 10^{27.7}$ W Hz$^{-1}$. The lower part of the table show the statistics for the radio galaxies and quasars with $P_{178} \ge 10^{27.7}$ W Hz$^{-1}$. The numbers of sources are shown in brackets.} 
\begin{center} 
\begin{tabular}{cllc}  
\emph{FRII type} & \multicolumn{2}{c}{\emph{Mean linear size} (kpc)} &\emph{Asymmetry}   \\ 
RG/QSO  & $-$FRII     & +FRII      &$\varepsilon$      \\ \hline \hline
RG      & 300  (9)    & 450  (10)  & $-0.8 \pm 0.8$            \\ \hline \hline  
RG+QSO  & 290  (27)   & 220  (54)  & $\sim0$           \\ 
RG      & 290  (19)   & 280  (30)  & $-0.27$           \\ 
QSO     & 280   (8)   & 140  (24)  &  +0.33                 
\end{tabular}
\end{center}
\end{table}

\begin{table}
\caption{The high luminosity sample divided into equal samples. The first sample has $10^{27.7}{\rm W}{\rm Hz}^{-1}<P_{178}<10^{28.7}{\rm W}{\rm Hz}^{-1}$ and the second $P_{178}>10^{28.7}{\rm W}{\rm Hz}^{-1}$.} 
\begin{center} 
\begin{tabular}{cllc}   
\emph{FRII type} &\multicolumn{2}{c}{\emph{Mean linear size} (kpc)} &\emph{Asymmetry}       \\ 
RG/QSO  & $-$FRII    & +FRII       & $\varepsilon$   \\ \hline \hline
RG+QSO  & 300  (16)  & 310  (24)   & $-0.3 \pm 0.4$          \\ 
RG      & 330  (13)  & 300  (18)   & $-0.44 \pm 0.52$         \\ 
QSO     & 400   (3)  & 230   (6)   & $\sim0 \pm 0.7$         \\ \hline \hline
RG+QSO  & 250  (11)  & 160  (30)   & $+0.27 \pm 0.26$           \\ 
RG      & 280  (6)   & 220  (12)   & $\sim 0 \pm 0.5$         \\ 
QSO     & 210  (5)   & 120  (18)   & $+0.44 \pm 0.28$            
\end{tabular}
\end{center}
\end{table}
\begin{table}
\caption{The mean projected linear sizes of narrow-line (NL), broad-line (BL) objects and low-excitation (LE) radio galaxies. The numbers of sources are in brackets.} 
\begin{center} 
\begin{tabular}{ccllc}  
\emph{Spectral type} & \multicolumn{3}{c}{\emph{Mean projected linear size} (kpc)} &\emph{Asymmetry}   \\ 
NL/BL/LE  & FRII   & $-$FRII     & +FRII     &$\varepsilon$    \\ \hline \hline
NL        & 310 (51) & 280 (18) & 320 (33)& $-0.1 \pm 0.32$             \\  
BL        & 220 (36) & 320 (10) & 180 (26)& $+0.23 \pm 0.28$            \\ 
LE        & 270 (11) & 300 (6)  & 240 (5) & $\approx -1 \pm 1$              
\end{tabular}
\end{center}
\end{table}

\subsection{Spectral classification} 

The asymmetry parameters and mean linear sizes for broad-line radio galaxies and quasars, narrow-line radio galaxies and low-excitation radio galaxies have been evaluated for both the $-$FRII and +FRII sources (Table 4). The same three giant sources were excluded from the sample.   For broad-line objects, the mean projected size of $-$FRII is larger than the mean projected size of +FRII sources ($320 > 180$ kpc), while for narrow-line galaxies the mean sizes are similar $(280 \sim 320\, {\rm kpc})$, consistent with orientation-based unified schemes. This result is not particularly surprising since the broad-line radio galaxies are naturally identified with the low redshift `quasar' population. 

There are 11 low-excitation radio galaxies in the sample, most of which are relatively low-luminosity sources $(P_{178}<10^{28.5}{\rm W}{\rm Hz}^{-1})$. The mean sizes of low-excitation and high-excitation radio galaxies are 240 kpc and 380 kpc respectively. The same result was obtained by Hardcastle {\it et al.} (1997) for a combined sample of powerful radio galaxies with $z<0.3$. The mean projected size of low-excitation radio galaxies is smaller than that of the narrow-line radio galaxies of the same luminosities.  The roughly equal numbers of the $-$FRII and +FRII low-excitation radio galaxies (Table 4) indicates that their structural asymmetry is caused almost entirely by intrinsic asymmetries, which implies non-relativistic expansion speeds $(v_0\ll {c})$ and/or preferential orientation of their axes close to the plane of the sky ($\theta\sim\pi/2$).  This supports the view that the low-excitation radio galaxies are a separate group of FRII sources with non-relativistic expansion velocities and that they are not unified with high-excitation sources (Laing {\it et al.} 1994).

\section{Numerical simulations}  

The expected distribution of the fractional separation differences for the radio galaxies and quasars have been modelled, with the aim of estimating quantitatively the mean speed of the lobes, its standard deviation, the typical intrinisic/environmental asymmetries of the FRII sources and the critical angle separating the radio galaxies and the quasars according to the orientiation-based unification scenario.

\subsection{Expansion speeds and intrinsic/environmental asymmetries}

A lower limit to the lobe advance speeds can be found from the expression (15). Adopting $\overline{x}_{\rm l} = 0.07 \pm 0.018$, we find
\begin{equation}
\overline{v}_{\rm l}/c \ga 2\overline{x}_{\rm l} = 0.14 \pm 0.036
\end{equation}
Adopting the simple Gaussian assumptions made in Section 3, a procedure has been developed for determining the best fitting values of the mean lobe speed standard, $\overline{v}_{\rm l}$, the standard deviation of the lobe speeds $\sigma_{v_{\rm l}}$ and the standard deviation of the disturbed velocities $\sigma_{v_{\rm d}}$. For a given set of parameters, 10,000 random values of these velocities were selected from the joint Gaussian distributions for each source component and the value of $\cos\theta$ were selected at random from a uniform distribution between 0 and 1. The best fitting values were found by minimising the parameter $M = (\sum |z_{\rm obs}(x_{\rm l}) - z_{\rm pr}(x_{\rm l})|)/N$; in this expression $z(x_{\rm l})\,{\rm d}x_{\rm l} = N(x_{\rm l})/N$, where $N(x_{\rm l})$ represents the observed (obs) or predicted (pr) numbers of sources in intervals of 0.1 in $x_{\rm l}$ and $N$ is the total number of sources.
\begin{figure}
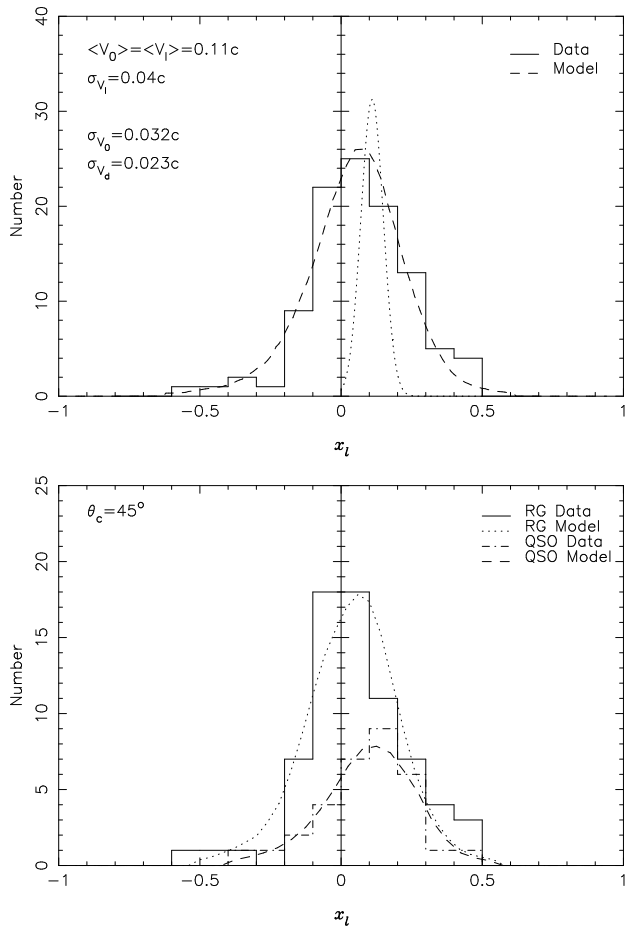

\begin{center}
\epsfig{file=TigFig3.ps, height=8.2cm, width=6.0cm, angle=-90.0}
\end{center}
\begin{center}
\epsfig{file=TigFig2.ps, height=8.2cm, width=6.0cm, angle=-90.0}
\end{center}
\caption{(a) Predicted distribution functions of the fractional separation differences $x_{\rm l}$ for the complete sample of radio sources compared with the observed distributions for the best-fitting values of the free parameters, $\overline{v}_0 = 0.11c$, $\sigma_{v_0} = 0.032c$,  $\sigma_{v_{\rm d}} = 0.023c$   and $\sigma_{v_{\rm l}} = 0.04c$. The dotted line shows the distribution function of the lobe speeds with $\overline v_0 = 0.11c$ and $\sigma_{\rm v_{\rm l}} = 0.04c$.  (b) The same as (a) for radio galaxies (dotted line) and quasars (dashed line) separately according to the unified scheme, compared with the observed distributions of $x_{\rm l}$ for radio galaxies (full line) and quasars (dot-dash line) for the same best-fitting values of the parameters.}\label{fig-9}
\end{figure}  

The two-sigma lower limit for the asymmetry parameter and the mean lobe speed correspond to $M\approx 0.17$ and the ranges of values of these parameters with $M < 0.17$  are $-0.47 < \varepsilon < 0.07 $ and $0.07 < \overline{v}_{\rm l}/c < 0.15$.  The best-fitting values of the parameters are:
$\overline{v}_{\rm l}/c = 0.11 \pm 0.013$,  $\sigma_{v_{\rm 0}} = (0.032 \pm 0.002)c$, $\sigma_{v_{\rm d}} = (0.023 \pm 0.002)c$  and $\sigma_{v_{\rm l}} = (0.04 \pm 0.004)c$. A comparison of the predictions of the model with these best-fitting values of the parameter with the observations is shown in Fig.~9(a).

The mean expansion speed of the lobes is comparable to the results of spectral ageing analyses, $\overline{v} = (0.13 \pm 0.08)c$, and, as expected, less than the mean speeds estimated for the symmetric model ($> 0.2c$). 

\subsection{Unified scheme}

We use (11) to model the fractional separation differences for radio galaxies and quasars separately. For illustrative purposes, the best-fitting values of the distributions found in section 7.1 have been used, that is, normal distributions for $v_0$, $v_{\rm jd}$ and $v_{\rm cjd}$ for both radio galaxies and quasars, the only free parameter being the critical angle $\theta_{\rm c}$. Values of $\cos\theta$ were selected at random from the uniform distribution from 1 to $\cos\theta_{\rm c}$ for quasars and from $\cos\theta_{\rm c}$ to 0 for radio galaxies.  The results of varying $\theta_{\rm c}$ are presented in Table 5. It can be seen that good agreement between predicted and observed asymmetry parameters, $\varepsilon_{\rm G} = -0.3$ and $\varepsilon_{\rm Q} = 0.33$, is found for $\theta_{\rm c}\sim 45^{\circ}$; furthermore, the predicted distribution functions of $x_{\rm l}$ are separately in good agreement with the distributions for galaxies and quasars separately, as can be seen in Fig.~9(b), and with the mean observed values of $\overline{x}_{\rm G}$ and $\overline{x}_{\rm Q}$, namely, $0.056 \pm 0.022$ and $0.102 \pm 0.028$ respectively (see Table 5).  

We conclude that the asymmetric relativistic model is in agreement with a unified scheme for $\theta_{\rm c} \sim 45^{\circ}$, the value favoured by Barthel (1989). The success of the modelling also indirectly confirms the validity of the estimates of the mean lobe speed and its standard deviation.

\begin{table}
\caption{Predicted asymmetry parameters and mean values of fractional separation differences for radio galaxies and quasars in a unified scheme.}

\begin{center}

\begin{tabular}{ccccc}    
$\theta_{\rm c}^{\circ}$ & $\varepsilon_{\rm\small{G}}$ & $\varepsilon_{\rm\small{Q}}$ & $\overline{x}_{\rm G}$ & $\overline{x}_{\rm Q}$\\ \hline
15 & $-0.12$ & 0.37 & 0.051 & 0.098\\
25 & $-0.15$ & 0.35 & 0.047 & 0.098\\
35 & $-0.22$ & 0.32 & 0.045 & 0.087\\
45 & $-0.28$ & 0.28 & 0.038 & 0.087\\ 
55 & $-0.38$ & 0.22 & 0.031 & 0.082\\
65 & $-0.52$ & 0.16 & 0.023 & 0.080\\ 
75 & $-0.67$ & 0.07 & 0.006 & 0.069
\end{tabular}

\end{center}

\end{table}

\section{Conclusions and discussion}

The principal conclusions of this analysis are as follows:
\begin{enumerate}

\item A statistical analysis of the sample of 3CRR sources shows that (a)  roughly one third of the FRII sources have a jet on the short lobe side; (b) the predictions of the asymmetric relativistic model can provide a good description of the structural properties of the radio sources in the sample; (c) the contributions of relativistic effects and intrinsic/environmental asymmetries to the structural asymmetry are of comparable importance. Relativistic effects are more important for quasars and high-luminosities, while intrinsic asymmetries are more important for low-luminosities and radio galaxies. 
  
\item Analysis of the correlations of the fractional separation difference with radio luminosity and linear size shows, that
(a) there is a trend for the expansion speeds of the lobes and intrinsic/environmental asymmetries to increase with increasing radio luminosity, and (b) intrinsic/environmental asymmetries are more important on small physical scales but they become less significant on larger physical scales. 

\item The mean lobe speed of FRII sources and its standard deviation, $\overline{v}_{\rm l}\sim(0.11\pm0.013)\,c$ and $\sigma_{v_{\rm l}} = 0.04$, have been determined from numerical simulations of the observed distribution function of fractional separation differences for both radio galaxies and quasars.  These speeds are consistent with ages determined by synchrotron ageing arguments.

\item The asymmetric relativistic model is in agreement with an orientation-based unified scheme in which the critical angle separating radio galaxies from quasars is about $45^{\circ}$. 

\end{enumerate}

The tendency for the intrinsic/environmental asymmetry effects to decrease at low-luminosities and large physical sizes is a direct reflection of the negative correlation between luminosity and linear size. The decrease of intrinsic asymmetries on large physical scales can be naturally explained if the intergalactic gas is more uniform at large distances from the active nuclei. However, an important question is whether environmental effects are responsible for all the observed asymmetries, or whether intrinsic jet asymmetries also need to be taken into account.   

Some observations suggest that intrinsically the jets may well be symmetric. In the Compact Symmetric Objects (CSO), two hot-spots are located symmetrically with respect to the active galactic nuclei and these are thought to be formed by  symmetric jets. Readhead {\it et al.} (1996) and Conway \& Owsianik (1998) have shown that CSOs are very young objects with ages of only  300 -- 3000 years and represent a very early phase in the evolution of classical double FRII radio sources. Saripalli {\it et al.} (1997) have detected two parsec scale jets in the giant double radio source DA240 with a flux density ratio of unity. This is the clearest instance of parsec scale jets of equal strengths in FRII sources.  

McCarthy {\it et al.} (1991) quantified the degree of asymmetry in the extended emission-line regions by the parameter $R_{\mbox{\scriptsize[OII]}}$, which is the ratio of the emission-line brightness on either side of the nucleus. They showed that the brighter emission-line region lies on the short lobe side in essentially all cases and considered this as evidence that environmental effects determine the fractional separation differences.  In this case, the large emission-line asymmetry parameters should correspond on average to those sources in which the intrinsic/environmental asymmetries dominate, that is, those FRII sources with $x_{\rm l} < 0$ and $x_{\rm l} \ga 0.26$. For the 21 FRII sources which are common to both the sample of McCarthy {\it et al.} (1991) and our own, the fractional separation differences are plotted against the emission-line asymmetry parameter in Fig.~10. Excluding the source 3C244.1 which has an extremely large value of $R_{\mbox{\scriptsize[OII]}}>20$, the mean emission-line asymmetry parameters, are found to be $>4.4$, $<2.5$ and $>3.9$, for three parameter regions $x_{\rm l} < 0$, $0 <x_{\rm l} \la0.26$ and $x_{\rm l} > 0.26$, respectively. It is noteworthy that the mean values of $R_{\mbox{\scriptsize[OII]}}$ in the first and third regions are larger than in the middle one, which illustrates both the important r\^ole of environmental asymmetries and the validity of the asymmetric relativistic model. 

\begin{figure}
\begin{center}
\epsfig{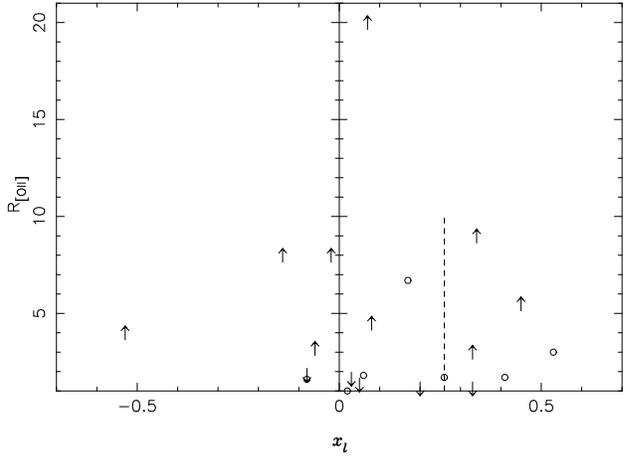}
\end{center}
\caption{The emission-line asymmetry parameter against fractional separation difference of radio lobes for 21 FRII sources. The upper and lower limits of $R_{\mbox{\scriptsize[OII]}}$ are denoted by $\downarrow$ and $\uparrow$, respectively.}\label{fig-10}

\end{figure} 

The magnitude of the intrinsic/environmental asymmetry is quite small relative to the true mean space velocities of advance of the lobes.  Adopting the standard deviation of 0.023$c$ for $\sigma_{v_{\rm d}}$, this amounts to about 20\% of the mean advance speed of the lobes. This is not particularly unexpected since the remarkable feature of the FRII sources is the fact that they are as symmetric as they are.  This small fractional velocity difference occurs, despite the large differences in the [OII] emission on either side of the radio source.  The particle densities in the emission line regions should be $\sim 10^2$ cm$^{-3}$, compared with typical densities of $\sim 10^{-2}$ cm$^{-3}$ for the interstellar and intergalactic environments of FRII radio sources. If the power supplied by the radio jet is assumed to be constant, the rate of advance of the front edge of the radio lobe is given by the balance of the ram pressure of the jet and internal pressure of the hot-spots $p \approx \rho v^2$.  Keeping $p$ constant, it can be seen that the rate of advance of the jet would be much less if it had to penetrate clouds with densities $10^4$ times greater than the intervening medium.  The answer almost certainly lies in the fact that the emission line gas has a very small filling factor, $\sim 10^{-6}$, and so there is only a small, but significant, probability of the jet encountering a region of high density gas.  Detailed modelling needs to be carried out to investigate the implications of the small value of $\sigma_{v_{\rm d}}/v_0$ for the properties of the [OII] clouds.   

\subsection*{Acknowledgements}

We are very grateful to Dr. Julia Riley for valuable discussions and for kindly supplying up-to-date information about 3CRR sample, and to Drs. Philip Best and Peter Scheuer for valuable comments.  We are also very grateful to Dr. Simon Garrington  for careful reading of the text and helpful comments, which have significantly improved this paper.   TGA gratefully acknowledges the award of a Royal Society-NATO Post-doctoral Fellowship and a Royal Society ex-quota award.

\subsection*{APPENDIX: DETAIL AND DESCRIPTION OF THE SAMPLE}

Information about the jet-side was available from the literature for 103 FRII radio sources from the 3CRR complete sample defined by Laing {\it et al.} (1983). All 103 radio sources were of high radio luminosity $P_{178}\geq 2.0 \times 10^{26}\,{\rm W\,Hz}^{-1}$.  The structural asymmetry of the sources was defined by the \emph{fractional separation difference} of lobes $x_{\rm l} = (r_{\rm j}-r_{\rm cj})/(r_{\rm j}+r_{\rm cj})$, where $r_{\rm j}$ and $r_{\rm cj}$ are the angular distances from the core to the furthest end of the lobes on the jet and counterjet sides respectively. The fractional separation difference of hot-spots $v_{\rm h}$ is the ratio of the difference and the sum of the of the core--hot-spots angular distances on the jet and counterjet sides respectively. The value $x_{\rm l}$ is negative if a jet (or approaching) side is shorter than counterjet side and is positive if jet side is longer. 

For the most of the FRII sources, values of fractional separation difference of hot-spots given in Table 6 are in agreement with the values published by Best {\it et al.} (1995) and McCarthy {\it et al.} (1991) within about 5\%. Significant discrepancies for 13 sources (3C: 9, 13, 153, 175.1, 184, 205, 217, 226, 236, 324, 352, 368, 432) are generally the result of new deep VLA observations providing improved estimates of $x_{\rm h}$. 

The core--lobe angular distances for more than half the FRII sources are taken from the published data of H2, BHL, FBB, FBP and PRM (see Table 6). For the remaining sources, we have measured the distances from the radio images following the procedure of Scheuer (1995): in our analysis, angular distances from the core have been measured to the second or third contour of the radio isophots rather than the lowest. 

There are uncertainties in the core--lobe angular distances for 3 objects: 3C171, 3C215 and 3C309.1. The last two objects have a strong bent lobe structures. We included these three sources in the final sample measuring the fifth lowest contour.              
The data are presented in the Table 6 in the following format:
\begin{description}
 \item[Column 1] IAU name.
 \item[Column 2] Common name.
 \item[Column 3] The jet or bright jet side.
 \item[Column 4] Presence of jet evaluated from VLA/VLBI jet detections (J), depolarization (DM) and spectral index (SI) asymmetries. 
 \item[Column 5] Definite (D) or probable (P) jets.
 \item[Column 6] Fractional separation difference of hot-spots $x_{\rm h}$.  
 \item[Column 7] Fractional separation difference of the lobes, including an indication of whether the jet lobe size is larger ($x_{\rm l}\geq0$) or smaller ($x_{\rm l}\leq0$) than counterjet lobe size.
  \item[Column 8] Identification of the active galaxy with radio galaxy (G) or quasar (Q).
 \item[Column 9] Emission class of the active galaxy nuclei: narrow-line (N), broad-line (B) and low-excitation (L) radio galaxies.   
 \item[Column 10] Redshift of source.
 \item[Column 11] Logarithm of the radio luminosity at $178\,{\rm MHz}$.   
 \item[Column 12] Linear sizes in kiloparsec (kpc), calculated from the data of Laing \& Riley (personal communication).
 \item[Column 13] First and second references indicate the jet-side and radio maps of the FRII sources which were selected to calculate the fractional separation difference of lobes and hot-spots, respectively. Single reference means that the jet-side and radio maps were taken from one source.    
      
\end{description}

\begin{table*}

\caption{The sample of 103 FRII radio sources}
\begin{tabular}{llcccrrccccrl}
 IAU name & Source & jet-side & J/DM/SI & D/P & $x_{\rm h}$ & $x_{\rm l}$ & 
 G/Q & Emission & z & ${\rm Log}\,P_{178}$ & Size & Reference \\
       &         &         &         &       &     &               &
       & class   &  & $({\rm W\,Hz^{-1}})$ & (kpc)&     \\ \hline\hline 
0007+124& 4C12.03& NE& J & PP  &  0.22&  0.21& G& E & 0.156&    27.1&  563& H2,LP1      \\  
0017+154& 3C9    & SE& J & DD  &  0.21&  0.19& Q& B & 2.012&    29.8&  115& BHL         \\  
0033+183& 3C14   & SE& J & D   &  0.21&  0.21& Q& B & 1.469&    29.1&  205& GCL         \\  
0040+517& 3C20   & WN& J & DP  &  0.02& -0.03& G& N & 0.174&    27.8&  202& H1,H2       \\  
0048+509& 3C22   & WN& J & D   &  0.12&  0.13& Q& N & 0.937&    28.8&  206& FBB         \\  
0106+729& 3C33.1 & SW& J & D   &  0.26&  0.27& G& B & 0.181&    27.3&  892& H2,vBJ      \\  
0107+315& 3C34   & EN& J & P   &  0.05&  0.02& G& N & 0.690&    28.5&  381& FBP,JLG     \\  
0123+329& 3C41   & SE& J & D   & -0.06& -0.06& G& N & 0.794&    28.5&  190& GCL,L89     \\  
0125+287& 3C42   & NW& J & P   & -0.03& -0.03& G& N & 0.395&    28.0&  179& FBP         \\  
0133+207& 3C47   & SW& J & D   &  0.10&  0.09& Q& B & 0.425&    28.4&  457& BHL         \\  
0138+136& 3C49   & WN& J & P   & -0.40& -0.32& G& N & 0.621&    28.3&    8& SSL         \\  
0154+286& 3C55   & ES& J & P   & -0.01& -0.02& G& N & 0.735&    28.9&  573& FBB         \\  
0210+860& 3C61.1 & NE& J & P   &  0.10&  0.10& G& N & 0.186&    27.7&  741& H2,LP1      \\  
0221+276& 3C67   & NW& DM&     &  0.42&  0.34& G& B & 0.310&    27.7&   13& GA,SSL      \\  
0229+341& 3C68.1 & NW& J & DD  & -0.10& -0.11& Q& B?& 1.238&    29.0&  448& BHL         \\ 
0356+102& 3C98   & NE& J & D   & -0.08& -0.13& G& N & 0.031&    26.3&  262& H2,LBD      \\  
0410+110& 3C109  & SE& J & D   &  0.02&  0.03& G& B & 0.306&    28.0&  500& H2,GFV      \\  
0411+141& 4C14.11& SE& J & D   & -0.23& -0.01& G& E & 0.207&    27.4&  381& H2,H1       \\  
0453+227& 3C132  & ES& J & P   & -0.02& -0.03& G& E & 0.214&    27.5&  102& H2,H1       \\  
0605+480& 3C153  & WS& J & PP  &  0.33&  0.05& G& N & 0.277&    27.8&   44& H2,H1       \\  
0651+542& 3C171  & ES& J & DD  &  0.03&  0.34& G& N & 0.238&    27.7&  171& H2,H1       \\  
0702+749& 3C173.1& NE& J & D   & -0.09& -0.10& G& E & 0.292&    27.8&  325& H2,H1       \\  
0710+118& 3C175  & WS& J & D   &  0.17&  0.19& Q& E & 0.768&    28.8&  391& BHL         \\  
0711+146& 3C175.1& ES& DM&     &  0.02&  0.00& G& N & 0.920&    28.8&   59& PRM,NRF     \\  
0725+147& 3C181  & ES& J & P   &  0.47&  0.37& Q& B & 1.382&    29.3&   49& MJF         \\  
0734+805& 3C184.1& NW& J & P   &  0.14&  0.13& G& N & 0.118&    26.9&  472& H2,LBD      \\  
0740+380& 3C186  & NW& J & D   &  0.10&  0.10& Q& B & 1.063&    29.1&   19& SMS         \\  
0745+560& DA240  & NE& J & DD  &  0.06&  0.03& G& E & 0.035&    26.1& 2073& SPP,KMS     \\  
0758+143& 3C190  & SW& J & D   &  0.20&  0.20& Q& B?& 1.197&    29.1&   58& SMS         \\  
0802+103& 3C191  & SE& J & D   &  0.29&  0.23& Q& B & 1.956&    29.6&   40& AG          \\  
0802+243& 3C192  & ES& J & P   & -0.08& -0.08& G& N & 0.060&    26.6&  308& H2,LBD      \\  
0809+483& 3C196  & SW& SI&     & -0.13&  0.11& Q& B & 0.871&    29.4&   84& L84,LM      \\  
0824+294& 3C200  & SE& J & D   &  0.29&  0.13& G& N?& 0.458&    28.1&  178& GCL,BBY     \\  
0832+143& 4C14.27& ES& J & P   & -0.20& -0.19& G& N & 0.392&    28.0&  216& LP1         \\  
0833+654& 3C204  & WN& J & D   & -0.17&  0.00& Q& B & 1.112&    29.0&  300& BHL         \\  
0835+580& 3C205  & SW& J & D   &  0.08&  0.16& Q& B & 1.534&    29.3&  154& LB          \\  
0838+133& 3C207  & ES& J & D   &  0.08&  0.09& Q& B & 0.684&    28.6&  111& BHB         \\  
0850+140& 3C208  & WS& J & D   & -0.13&  0.07& Q& B & 1.110&    29.1&   94& BHL         \\  
0855+143& 3C212  & NW& J & D   &  0.06&  0.07& Q& B?& 1.049&    29.0&   77& BP,L88      \\  
0903+169& 3C215  & ES& J & DD  & -0.36&  0.09& Q& B & 0.411&    28.0&  384& BHL         \\  
0905+380& 3C217  & WN& J & P   &  0.33&  0.29& G& N & 0.897&    28.7&  101& PRM,BLR     \\  
0917+458& 3C219  & SW& J & DP  &  0.00& -0.03& G& B & 0.174&    27.8&  723& H2,CBB      \\  
0926+793& 3C220.1& EN& J & D   & -0.12& -0.08& G& N & 0.610&    28.5&  230& BBY         \\  
0936+361& 3C223  & NW& J & PP  & -0.02&  0.01& G& N & 0.137&    27.1&  962& H2,LBD      \\  
0941+100& 3C226  & NW& J & P   &  0.06&  0.06& G& N & 0.818&    28.8&  289& BHB         \\  
0947+145& 3C228  & SW& J & D   & -0.07& -0.07& G& N & 0.552&    28.6&  333& JLG,BBY     \\  
0958+290& 3C234  & EN& J & D   &  0.15&  0.15& G& N & 0.185&    27.7&  439& H2,H1       \\  
1003+351& 3C236  & EN& J & D   &  0.32&  0.19& G& E & 0.099&    26.8& 5946& BMJ         \\  
1008+467& 3C239  & WS& DM&     &  0.29&  0.12& G& N?& 1.790&    29.5&   94& LP2,LPR     \\  
1009+748& 4C74.16& SW& J & D   & -0.23& -0.14& G& - & 0.810&    28.6&  330& GCL,LRL     \\  
1030+585& 3C244.1& SE& J & P   &  0.07&  0.05& G& N & 0.428&    28.3&  352& FBP         \\  
1040+123& 3C245  & WN& J & D   &  0.23&  0.19& Q& B & 1.029&    28.9&   42& L89,ALB     \\  
1056+432& 3C247  & WS& DM&     &  0.17&  0.14& G& N & 0.750&    28.5&  105& LP2,LPR     \\  
1100+772& 3C249.1& ES& J & DD  & -0.34&  0.23& Q& B & 0.311&    27.7&  247& BHL         \\  
1111+408& 3C254  & WN& DM&     &  0.74&  0.44& Q& B & 0.734&    28.8&  106& LP2,LPR     \\  
1137+660& 3C263  & ES& J & D   & -0.26& -0.32& Q& B & 0.656&    28.5&  346& BHL         \\  
1140+223& 3C263.1& SW& DM&     &  0.05&  0.23& G& N & 0.824&    28.8&   56& LP2,LPR     \\  
1142+318& 3C265  & WN& J & P   &  0.20&  0.20& G& N & 0.811&    28.9&  643& FBB         \\  
1143+500& 3C266  & NW& SI&     &  0.05& -0.04& G& N & 1.275&    29.1&   37& LPR         \\  
1147+130& 3C267  & WS& SI&     & -0.06& -0.07& G& N & 1.142&    29.1&  327& PRM         \\  
1157+732& 3C268.1& WN& DM&     &  0.11&  0.12& G& N & 0.974&    29.0&  391& PRM         \\  
1206+439& 3C268.4& SW& J & D   & -0.06& -0.06& Q& B & 1.400&    29.1&   94& BBY         \\  
1218+339& 3C270.1& SW& J & D   & -0.09&  0.00& Q& B & 1.519&    29.2&  102& ALB,SBC     \\  
1241+166& 3C275.1& NW& J & D   &  0.21&  0.28& Q& B & 0.557&    28.5&  133& SBC         
 
\end{tabular}
\end{table*}

\begin{table*}
\contcaption{}
\begin{tabular}{llcccrrccccrl}
 IAU name & Source & jet-side & J/DM/SI & D/P & $x_{\rm h}$ & $x_{\rm l}$ & 
 G/Q & Emission & z & ${\rm Log}\,P_{178}$ & Size & Reference \\
       &         &         &         &       &     &               &
       & class   &  & $({\rm W\,Hz^{-1}})$ & (kpc)&     \\ \hline\hline 

1251+159& 3C277.2& EN& DM&     &  0.45&  0.43& G& N & 0.766&    28.6&  473& PRM         \\  
1254+476& 3C280  & WS& DM&     &  0.08&  0.10& G& N & 0.996&    29.1&  123& LP2,LPR     \\  
1319+428& 3C285  & EN& J & DP  & -0.08& -0.08& G& E & 0.079&    26.5&  271& H2,vBD      \\  
1343+500& 3C289  & WN& DM&     &  0.01&  0.01& G& N & 0.967&    28.8&   85& LP2,BLR     \\  
1404+344& 3C294  & WS& J & D   &  0.19&  0.18& G& N & 1.780&    29.4&  126& MSD         \\  
1409+524& 3C295  & SE& SI&     &  0.17&  0.20& G& N & 0.461&    28.9&   36& PT          \\  
1419+419& 3C299  & WS& DM&     &  0.53&  0.44& G& N & 0.367&    27.9&   71& LP2,LPR     \\  
1420+198& 3C300  & WN& J & D   &  0.41&  0.38& G& N & 0.270&    27.8&  516& H2,H1       \\  
1441+522& 3C303  & WN& J & D   &  0.00&  0.14& G& B & 0.141&    27.0&  140& H2,LP1      \\  
1458+718& 3C309.1& ES& J & D   & -0.13& -0.22& Q& B & 0.904&    28.9&   24& KWP         \\  
1517+204& 3C318  & SW& J & PP  &  0.43&  0.49& G& E & 0.752&    28.6&    6& SMS         \\  
1529+242& 3C321  & ES& J & D   &  0.06&  0.10& G& N & 0.096&    26.8&  689& LW          \\  
1547+215& 3C324  & EN& J & P   &  0.07&  0.06& G& N & 1.206&    29.1&   86& FBB,BCG     \\  
1549+202& 3C326  & WS& J & D   &  0.45&  0.28& G& B & 0.090&    26.9& 2664& KMS         \\ 
1549+628& 3C325  & WN& J & P   &  0.14&  0.13& Q& N & 0.860&    28.8&  133& FBP         \\  
1618+177& 3C334  & SE& J & DD  & -0.21& -0.05& Q& B & 0.555&    28.3&  430& BHL         \\ 
1622+238& 3C336  & SW& J & DD  & -0.33& -0.17& Q& B & 0.927&    28.7&  183& BHL         \\  
1626+278& 3C341  & WS& J & DD  & -0.14& -0.12& G& N & 0.448&    28.1&  475& BP,LP1      \\ 
1627+234& 3C340  & WS& J & D   & -0.02& -0.03& G& N & 0.754&    28.5&  382& JLG         \\  
1627+444& 3C337  & WN& DM&     &  0.33&  0.28& G& N & 0.630&    28.4&  332& PRM         \\  
1658+471& 3C349  & SE& J & P   &  0.03&  0.03& G& N & 0.205&    27.4&  352& H2,H1       \\ 
1704+608& 3C351  & EN& J & DD  & -0.18& -0.01& Q& B & 0.371&    28.0&  402& BHL         \\ 
1709+460& 3C352  & NW& J & D   &  0.03&  0.03& G& N & 0.806&    28.6&  115& GCL,BLR     \\  
1723+510& 3C356  & NW& DM&     &  0.34&  0.32& G& N & 1.079&    28.9&  643& PRM,FBB     \\  
1732+160& 4C16.49& SE& J & D   &  0.24&  0.25& Q& B & 1.296&    29.1&  138& GCL,BMS     \\  
1759+137& 4C13.66& NW& SI&     & -0.65& -0.59& G& - & 1.500&    29.2&   51& RLL         \\  
1802+110& 3C368  & SN& J & D   & -0.16& -0.16& G& N & 1.132&    29.1&   68& BCG         \\ 
1833+326& 3C382  & EN& J & D   &  0.02& -0.02& G& B & 0.058&    26.5&  288& H2,BBL      \\
1842+455& 3C388  & WS& J & DP  &  0.01& -0.11& G& E & 0.091&    27.0&   97& H2,BCH      \\
1845+797& 3C390.3& NW& J & DP  &  0.18&  0.18& G& B & 0.057&    26.9&  323& LP3         \\ 
1939+605& 3C401  & SW& J & D   &  0.17&  0.09& G& E & 0.201&    27.6&  102& H2,H1       \\  
2120+168& 3C432  & SE& J & D   &  0.19&  0.12& Q& B & 1.805&    29.4&  109& BHL         \\  
2141+279& 3C436  & SE& J & D   & -0.14& -0.09& G& N & 0.214&    27.6&  461& H2,H1       \\  
2153+377& 3C438  & NW& J & DD  &  0.12&  0.05& G& E & 0.290&    28.3&  119& H2,H1       \\  
2203+292& 3C441  & NW& J & D   & -0.53& -0.41& G& N & 0.707&    28.5&  260& FBP         \\  
2243+394& 3C452  & EN& J & DD  & -0.02& -0.02& G& N & 0.081&    27.2&  514& BBL         \\  
2252+129& 3C455  & WS& J & D   &  0.42&  0.26& Q& B & 0.543&    28.3&   24& BHB         \\  
2309+184& 3C457  & NE& J & P   & -0.04&  0.03& G& N & 0.427&    28.1& 1261& LP1         \\ 
2352+796& 3C469.1& NW& DM&     & -0.05& -0.03& G& E & 1.336&    29.1&  637& PRM          
   
\end{tabular}    
 
\end{table*}

\label{lastpage}

\end{document}